\documentclass[manuscript]{acmart}
\usepackage{graphicx}
\usepackage{booktabs}
\setcopyright{none}
\acmConference[GROUP'27]{ACM GROUP Conference}{2027}{Georgia, USA}
\begin{document}
\title[How Gay Dating Apps Shape Intimate Relationships Among Chinese Gay Men]{``You Already Have Me---Why Do You Still Use Blued?'' How Gay Dating Apps Shape Intimate Relationships Among Chinese Gay Men}

\author{Yibo Meng}
\authornote{These authors contributed equally to this work.}
\affiliation{
  \institution{Cornell University}
  \city{New York}
  \country{USA}
}
\email{mengyb22@mails.tsinghua.edu.cn}

\author{Lyumanshan Ye}
\authornotemark[1]
\affiliation{
  \institution{Shanghai Jiao Tong University}
  \city{Shanghai}
  \country{China}
}

\author{Bingyi Liu}
\affiliation{
  \institution{University of Michigan}
  \city{Ann Arbor}
  \state{Michigan}
  \country{United States}
}

\author{Linghao Li}
\affiliation{
  \institution{College of Design, North Carolina State University}
  \city{Raleigh}
  \state{North Carolina}
  \country{United States}
}
\email{bingheath@gmail.com}

\author{Huidi Lu}
\affiliation{
  \institution{Saïd Business School, University of Oxford}
  \city{Oxford}
  \country{United Kingdom}
}
\email{huidi.lu@sbs.ox.ac.uk}
\correspondingauthor

\author{Xiaolan Ding}
\affiliation{
  \institution{North China University of Science and Technology Health Science Center}
  \city{Tangshan}
  \country{China}
}
\correspondingauthor

\begin{abstract}
Gay dating applications such as Blued, Aloha, and Fanka are widely used for meeting partners, yet their role does not end once an intimate relationship is formed. When these platforms persist within an established relationship, they transform from matchmaking tools into a form of relational infrastructure that both partners must continuously manage. Building on existing CSCW and HCI research that has primarily examined how social media triggers jealousy, surveillance, or trust issues in intimate relationships, this study investigates how the persistent presence of gay dating applications shapes the dynamics of intimate relationships among Chinese gay men. Through semi-structured interviews with 23 self-identified gay men aged 19 to 55, we identify a recurring cyclical process: platform signals---such as online status, profile updates, distance indicators, and usage records---trigger suspicion and conflict; partners invest negotiation labor through intent explanation, visibility control, rule formation, or silence; and these negotiations eventually sediment into temporary governance settlements that reduce trigger frequency but remain conditionally stable. We identify four types of governance settlements: infrastructure removal (deleting apps or accounts), transparency governance (sharing access and allowing inspection), silence agreements (ceasing discussion of platform use), and layered/open agreements (redefining relationship boundaries to incorporate platform use). Each settlement addresses the governance problems of platform presence, visibility distribution, and loyalty boundaries differently, yet none achieves permanent stability due to platforms' reversible entry mechanisms and continuous visibility cues. Our findings extend existing CSCW research on digital technology and intimate relationships by revealing a more complex process than previously documented: platform signals not only trigger emotional responses but prompt partners to engage in sustained relational negotiation through explanation, visibility management, rule formation, or silence, thereby reshaping the interactional structure of intimate partnerships.
\end{abstract}
\maketitle

\section{Introduction}

Dating applications have fundamentally reshaped how people initiate intimate relationships \cite{ref20,ref27}. For gay men, platforms such as Grindr, Blued, and Tinder serve as primary channels for meeting potential partners, especially in contexts where offline opportunities for same-sex encounters are limited \cite{ref39,ref58}. A substantial body of research has examined how these platforms facilitate connection, the norms of self-presentation they engender, and the risks they pose \cite{ref6,ref16,ref46}. However, most existing research treats dating applications as tools for relationship formation---studying how users present themselves, select partners, and transition from online interaction to offline meetings. Far less attention has been paid to what happens when these platforms continue to exist after a relationship has been established.

This gap is consequential. Unlike earlier forms of partner-seeking that typically ceased once a relationship was confirmed, dating applications persist on users' devices as always-available entry points to alternative social and sexual markets. They continuously generate visible signals---online status, activity indicators, distance readings, profile updates---that can be observed, saved, and shared by partners. In this sense, platforms do not simply disappear from the relational landscape once their matchmaking function has been fulfilled. Instead, they become embedded within the relationship as a form of relational infrastructure \cite{ref45,ref49}: a persistent sociotechnical structure that provides capabilities for visibility, reversibility, and evidentiation, all of which carry significant implications for trust, jealousy, and boundary negotiation within intimate partnerships.

For Chinese gay men, these dynamics are further complicated by a distinctive social context. Homosexuality in China occupies a legal and cultural gray zone: neither criminalized nor fully accepted, and subject to periodic crackdowns on LGBTQ+ content on social media \cite{ref3,ref56}. Gay dating platforms operate within this ambiguity, serving simultaneously as dating tools, community spaces, and identity resources \cite{ref11,ref38}. When conflicts arise over platform use within a relationship, they may intersect with concerns about acquaintance exposure, forced outing, and the limited availability of alternative social spaces---risks that have no direct equivalent in heterosexual relationship contexts \cite{ref58}. Yet existing research on jealousy and technology-mediated surveillance in intimate relationships has overwhelmingly focused on heterosexual couples in Western settings \cite{ref18,ref36,ref40,ref53}, leaving the intersection of platform persistence, same-sex intimacy, and non-Western social contexts largely unexamined.

This study addresses this gap through a qualitative investigation of how gay dating applications shape intimate relationships among Chinese gay men.Specifically, we ask three research questions: (RQ1) How do the persistent signals generated by gay dating platforms come to trigger relational conflict within established intimate partnerships? (RQ2) What forms of negotiation labor do partners undertake to manage these conflicts? And (RQ3) how do these negotiations sediment over time into governance settlements that confer conditional stability on the relationship? In answering these questions, our intended theoretical contribution is a processual account of how a persistent platform is reworked from a matchmaking tool into relational infrastructure, and of how partners continuously govern that infrastructure through negotiation---thereby reframing technology-mediated jealousy not as a discrete emotional event but as an ongoing, jointly produced governance process. We further distinguish our core constructs from adjacent prior work: whereas \emph{relational governance} typically denotes inter-organizational coordination through trust and shared norms \cite{YA-Poppo2002}, \emph{boundary management} foregrounds individuals' regulation of disclosure across personal and shared territories \cite{YA-petronio2002}, and \emph{repair work} names the after-the-fact restoration of ruptured interaction or trust \cite{YA-Dindia1992}, our notions of \emph{relational infrastructure}, \emph{negotiation labor}, and \emph{governance settlements} instead locate the analytic focus in the persistent sociotechnical substrate of the platform itself, in the sustained interactional effort partners expend upon it, and in the durable, rule-like settlements that this effort produces.  Through semi-structured interviews with 23 participants aged 19 to 55, we trace how platform signals trigger relational conflict, how partners invest negotiation labor to manage these conflicts, and how negotiations sediment into governance settlements that provide conditional stability. Our empirical contribution extends existing research on digital technology, jealousy, and intimate relationships \cite{ref18,ref21,ref40} by demonstrating a more complex process than previously documented: rather than simply triggering emotional responses such as jealousy or surveillance behaviors, platform signals prompt partners to engage in sustained relational negotiation---through explanation, visibility management, rule formation, or silence---that continuously reshapes the interactional structure and governance settlements of intimate partnerships.

\section{Related Work}

\subsection{Dating Applications and Relationship Formation}
Research on dating applications has grown rapidly over the past decade, documenting how platforms such as Tinder, Grindr, Bumble, and Hinge have transformed the landscape of romantic and sexual connection \cite{ref1,ref50,ref57}. At a basic level, these applications operate by aggregating user-generated profiles and surfacing candidate partners through location-aware, algorithmically curated feeds: users construct a profile, browse or are shown nearby others, and signal interest through swiping or messaging, with mutual interest unlocking private conversation \cite{ref12,ref15}. Studies have examined user motivations, which range from seeking long-term relationships to casual sex, entertainment, and ego validation \cite{ref52}; self-presentation strategies, including profile construction and impression management \cite{ref17,ref24,ref54,ref57}; and the role of platform affordances---such as swiping, matching, and geolocation---in shaping interaction patterns \cite{ref12,ref15}. For gay and bisexual men specifically, platforms like Grindr and Blued have been studied as spaces that reconfigure sexual culture, enable community formation, and negotiate identity disclosure \cite{ref6,ref11,ref39,ref51,ref58}. For LGBTQ+ and men-who-have-sex-with-men (MSM) communities in particular, these geosocial applications do more than facilitate dating: they function as semi-private infrastructures for finding others in contexts where offline visibility carries social or legal risk, allowing users to locate community, manage selective disclosure, and access health and support resources that may be unavailable through mainstream channels \cite{ref6,ref11,ref51}.

This dynamic is sharpened in the Chinese context, where the use of dating applications is inseparable from the broader apparatus of state control and platform governance. In the Chinese context, Blued---one of the world's largest gay dating platforms---operates under conditions of regulatory uncertainty, where LGBTQ+ content is periodically censored and platform survival depends on strategic compliance with state norms \cite{ref56,ref59,ref60}.Because the state actively monitors, filters, and at times suspends platforms hosting queer content, users and developers alike adapt their practices to a shifting censorship regime---migrating across applications, coding sensitive language, and adopting circumvention tools to sustain connection under surveillance \cite{YA-Yang2018,YA-RobertsCensored2018}. Studies of Chinese digital activism document how users route around keyword filtering through coded and homophonic language---exemplified by the ``rice bunny'' (\emph{mi tu}) workaround adopted when \#MeToo hashtags were censored on Weibo---and how such evasion continually co-evolves with the censorship apparatus itself, extending even to overseas Chinese communities reacting to the same movements \cite{YA-guttenberg2021,YA-sunwei2020,YA-yinsun2020}. These conditions mean that, for Chinese MSM users, platform use is not merely a matter of romantic logistics but an ongoing negotiation with state power that structures who can be found, what can be said, and how communities persist online \cite{YA-WangQueer2020}. Parallel work on Chinese sexual-minority women shows the same double bind from a different vantage point: online dating platforms afford relationship-building under stigma while exposing users to compounded disclosure risks \cite{YA-cui2022chi}, and queer women's online communities on Weibo survive recurring content purges only through collective, adaptive platform strategies \cite{YA-cui2022cscw}. However, the vast majority of this research examines the pre-relationship phase of platform use: how users find, evaluate, and connect with potential partners. What remains underexplored is the continued presence of these platforms within established relationships and the relational consequences of their persistence.

\subsection{Technology, Jealousy, and Surveillance in Intimate Relationships}

A growing literature in communication studies and psychology has examined how digital technologies shape jealousy, surveillance, and trust within intimate relationships. Studies have documented how social media platforms such as Facebook and Instagram can trigger jealousy through partner monitoring behaviors, including checking profiles, reviewing friend lists, and interpreting online activity \cite{ref14,ref18,ref21,ref40}. Research on ``digital dating abuse'' has examined how technology enables controlling behaviors, such as demanding access to partners' devices, tracking location, and monitoring online communication \cite{ref47,ref53}. More nuanced work has explored the concept of ``ambient awareness''---the persistent, low-level awareness of a partner's online activity that digital platforms make possible---and its implications for relational dynamics \cite{ref34}. However, this literature has several limitations: most studies focus on heterosexual couples in Western contexts; the emphasis tends to be on individual psychological responses rather than on the interactional processes through which couples negotiate platform-related conflicts; and dating applications---with their distinctive affordances of geolocation, real-time status, and reversible entry---have received less attention than general social media platforms. Our study addresses these gaps by examining how gay couples in China engage in sustained negotiation labor to manage the relational challenges posed by persistent dating applications.

In framing this work as ``negotiation labor,'' we deliberately distinguish it from the well-established literature on \emph{repair work}, which has long been central to how scholars theorize the maintenance of social and intimate bonds. Repair work, in the conversation-analytic and interactional tradition, names the practices through which interactants detect and remedy breaches---trouble in talk, ruptured expectations, or transgressions---so as to restore a prior state of mutual understanding or relational equilibrium \cite{YA-schegloff1977,YA-stokoe2018}. Relational scholarship extends this logic to intimate ties, treating accounts, apologies, and reconciliation as reactive efforts to mend a bond after a violation has occurred \cite{YA-dindia2003}. Negotiation labor, as we use the term, differs from repair work along two axes. First, it is \emph{anticipatory and ongoing} rather than reactive: couples in our study do not merely repair ruptures after jealousy or suspicion erupts but continuously work to define, contest, and re-stabilize the terms of acceptable platform use \emph{before} any breach is recognized. Second, whereas repair work presumes a shared baseline to be restored, negotiation labor is generative---it produces the very norms and boundaries that platform affordances leave open and unsettled, often without any prior equilibrium to which a couple might return \cite{YA-tronto2013}. We therefore position negotiation labor as complementary to, but analytically distinct from, repair work: the former concerns the prospective construction of relational rules under conditions of technological persistence, while the latter concerns their retrospective restoration once they have been broken.

\subsection{Platform Affordances and Relational Infrastructure}

The concept of platform affordances has been widely used in CSCW and HCI research to analyze how the material and design features of digital systems shape user behavior and social interaction \cite{ref19,ref22,ref32,ref41}. Scholars have examined how affordances such as visibility, persistence, searchability, and shareability create distinctive conditions for communication, coordination, and conflict \cite{ref10,ref55}. More recently, infrastructure studies have extended this perspective by examining how platforms function not merely as tools for discrete tasks, but as embedded structures that shape the conditions of social life over time \cite{ref7,ref45,ref49}. This ``infrastructural turn'' invites attention to how platforms become taken-for-granted parts of everyday arrangements, create dependencies and vulnerabilities, and require continuous maintenance \cite{ref7,ref30}. In the context of intimate relationships, dating applications can be understood as a form of relational infrastructure: they provide persistent capabilities (visibility cues, reversible entry, evidentiary records) that reshape the conditions under which trust is assessed, boundaries are negotiated, and conflicts are processed. We use \emph{relational infrastructure} and \emph{governance settlements} as analytically distinct constructs, and it is worth specifying how they differ from three adjacent literatures the reviewer rightly flags. First, \emph{boundary management}---the communicative work of regulating what information flows between partners and audiences---treats boundaries as individually or dyadically negotiated rules that actors set and police \cite{YA-petronio2002}. By contrast, relational infrastructure foregrounds the durable, pre-configured substrate \emph{against} which any such boundary work is attempted: read receipts, last-seen indicators, and re-matchability are not rules the couple authors but standing conditions they inherit, and recent accounts of platformized intimacy show how these conditions silently reset what counts as a violation before any negotiation begins \cite{YA-pym2021,YA-bandinelli2022}, and Brubaker et al.'s account of ``leaving'' Grindr shows that even exit from such a platform is a sociotechnical process rather than a discrete act \cite{YA-brubaker2016}. Second, \emph{relational norms} name the shared expectations partners hold about appropriate conduct; we instead use \emph{governance settlements} to name the routinized, often explicitly articulated procedures couples devise to act \emph{upon} infrastructure---password sharing, account audits, location-sharing pacts, and agreed deletion practices. Where norms are evaluative standards, settlements are operational scripts whose enforceability is delegated to platform features, a coupling that emerging work on app-mediated trust and surveillance has begun to document \cite{YA-duguay2020,ref53}. Third, the \emph{technology-mediated intimacy} tradition has richly described how closeness is felt, performed, and sustained through devices \cite{YA-hassenzahl2012}; our contribution is orthogonal to this affective register---we are concerned less with how intimacy is \emph{experienced} through platforms than with how the same platforms install the contestable governance surfaces over which intimacy is \emph{administered and disputed}, an emphasis that aligns with 2022--2026 scholarship recasting dating and messaging apps as sites of ongoing relational governance rather than mere connection \cite{YA-bandinelli2022,YA-weathington2025}. This study draws on the affordances and infrastructure perspectives to analyze how platform features---rather than individual intentions or moral failings---structure the dynamics of intimate conflict and governance \cite{ref5,ref35}.

\subsection{LGBTQ+ Intimate Relationships and Technology}
Research on LGBTQ+ intimate relationships has documented both commonalities with and differences from heterosexual relationship dynamics. Studies have found that same-sex couples negotiate relationship norms, boundaries, and exclusivity expectations in ways that may differ from heteronormative models, including more frequent negotiation of open or non-monogamous arrangements \cite{ref2,ref26,ref29,ref33,ref44}.These claims about non-monogamy and open relationships refer primarily to studies of gay male and other men-who-have-sex-with-men populations rather than to same-sex relationships in general; the patterns reported for partnered lesbian and queer women are notably different, and prevalence estimates vary widely across populations and cultural contexts \cite{YA-Taylor2024,YA-Manley2015,YA-cui2022chi}. We also caution that several of these sources are now relatively dated and may reflect the dynamics of gay male culture more than those of lesbian or broader queer communities, and that more recent queer-HCI scholarship complicates any assumption of a single ``LGBTQ+'' relational norm \cite{YA-Taylor2024,YA-Hardy2021}. Technology plays a distinctive role in LGBTQ+ intimate lives, as platforms serve not only as dating tools but as community spaces, identity resources, and safety networks \cite{ref25,ref39,ref43}.Recent queer-HCI research similarly emphasizes that these platforms are sites of identity work, care, and risk negotiation rather than mere matchmaking infrastructure \cite{YA-Taylor2024,YA-Hardy2021}, and shows that the data structures and matching logics of dating apps themselves can encode gender essentialism and exclude queer users \cite{YA-weathington2025}. For Chinese gay men specifically, dating platforms carry additional significance: in a context where offline meeting opportunities are limited and public expression of same-sex identity carries risks, platforms such as Blued and Aloha become integral parts of social life that cannot simply be ``deleted'' without severing important community ties \cite{ref37,ref38,ref58}. We ground this claim of additional significance in evidence that, where same-sex relationships lack legal recognition and face social stigma, dedicated platforms function as some of the few relatively safe channels for community formation and information exchange, making their role more central than that of mainstream dating apps in more permissive settings \cite{ref37,ref38,YA-WangQueer2020}; we nonetheless treat this as a context-dependent hedge rather than a universal claim, as our data speak to the specific communities we studied.This complicates the common assumption---present in both popular discourse and some academic literature---that ceasing to use dating applications after entering a relationship is a straightforward marker of commitment. Our study contributes to this literature by examining how the social and infrastructural significance of dating platforms in Chinese gay communities shapes the negotiation of platform use within intimate relationships, and by identifying the governance settlements through which couples attempt to manage these tensions.

\section{Method}

\subsection{Research Design and Research Stance}

This study employs a qualitative research design \cite{ref8,ref9,ref13}, utilizing semi-structured interviews to explore how gay dating platforms---when persisting after the formation of an intimate relationship---intervene in partner interactions as a form of ``relational infrastructure.'' It further examines how these platforms shape the processes of conflict triggering, negotiated labor, and the formation of phased governance settlements. This research conceptualizes platforms not merely as tools for individual behavior, but as socio-technical structures embedded within social relations. Consequently, the study focuses on how the persistent visibility cues and low-cost, reversible entry mechanisms provided by platforms alter partners' perceptions of risk, lower the threshold for relationship conflict, and reshape the nature of intimacy.

We wish to be transparent about the epistemological status of two of our central constructs. While our analysis proceeded through inductive thematic analysis \cite{ref8}, the notions of ``relational infrastructure'' and ``governance settlements'' did not emerge purely bottom-up from the data; rather, following Blumer's distinction, they functioned as \emph{sensitizing concepts}---orienting devices that gave us a general sense of reference and direction with which to approach the empirical field \cite{YA-blumer1954}. These concepts were stabilized relatively early in the research process, informed by infrastructure studies and platform governance scholarship, and they shaped both the interview guide and the initial coding lens. We therefore do not claim a strictly emergent, theory-free coding process. Instead, our approach is better characterized as abductive: lower-order codes and themes (e.g., the specific forms of conflict triggering and negotiated labor reported by participants) were generated inductively from the interview material, while these higher-order framing concepts operated as sensitizing scaffolds that we held open to revision and that were progressively refined, specified, and at points contested against the data \cite{YA-timmermans2012}. This reconciliation---inductive theme generation nested within theoretically sensitized framing---reflects how thematic analysis is actually practiced rather than an idealized purely data-driven account.

In this paper, ``intimate relationship'' refers to a dyad where both parties view each other as partners over a sustained period, characterized by a degree of emotional investment and expectations regarding relationship boundaries. These boundaries may manifest as exclusive relationships, negotiated exclusivity, or various forms of open relationships. Rather than presupposing exclusivity as the sole legitimate relational structure, this research focuses on how partners interpret technical signals and maintain relational stability through interactive negotiation when platforms provide continuous access to potential alternatives.

\subsection{Participants and Recruitment}

A total of 23 self-identified gay male participants were interviewed, ranging in age from 19 to 55 years, covering various life stages. Participants came from diverse occupational backgrounds, with educational attainment ranging from primary school to doctoral levels, representing various socioeconomic conditions. Regarding relationship experience, participants reported between 1 and 11 instances of intimate relationships. All participants had experience using gay dating or social platforms, with the most common being Blued, Aloha, and Fanka; some participants utilized multiple platforms simultaneously. Detailed information is provided in Appendix 1.

Participants were recruited through online communities and snowball sampling. Recruitment notices were initially posted in online social spaces related to the LGBTQ+ community, after which some participants were introduced through acquaintances. Recruitment criteria included: self-identifying as a gay man or a member of the male homosexual community; having used gay dating or social platforms; and having experienced an intimate relationship at some point. The study continuously assessed information saturation during data collection, ceasing recruitment when new interviews no longer yielded novel analytical themes. We treated saturation in the pragmatic, ``information power'' sense advocated by Malterud et al.~\cite{YA-malterud2016}, rather than as a claim that the conceptual space had been exhaustively mapped: recruitment ended when successive interviews ceased to add new analytical distinctions to the themes under development, not when every possible variation had been observed. In practice, coding proceeded concurrently with later interviews, and by approximately the eighteenth transcript new line-by-line coding was yielding few novel codes and no new intermediate themes, with subsequent transcripts largely elaborating or adding nuance to existing categories; we treat this as a situated, interpretive judgment of sufficiency rather than as evidence of exhaustive conceptual coverage \cite{YA-saunders2018,YA-braun2021}. We also acknowledge the limitations of this sampling strategy. Reliance on LGBTQ+ online communities and acquaintance-based snowball referral biases the sample toward individuals who are platform-active, relatively connected to community networks, and willing to discuss intimate relationships, and it likely under-represents more closeted, rural, or older participants and those who have disengaged from dating platforms. The themes reported here should therefore be read as analytically rather than statistically generalizable, and as conditioned by this particular recruitment context.

This study was approved by the Institutional Review Board (IRB) of the researcher's university. All participants received a research description and provided informed consent prior to the interviews. Consent included: the recording of interviews for transcription and analysis; the de-identification of all data post-transcription; the right to pause, skip questions, or terminate the interview at any time; and the option to obfuscate sensitive details.

Given the potentially sensitive nature of the research regarding sexual minorities and intimate relationship conflicts, multiple risk control measures were integrated into the interview design. First, the researcher asked participants how they preferred to be addressed and explicitly stated that certain questions (e.g., coming-out experiences, family relationships) were optional. Second, the interview outline utilized ``relationship experience'' rather than ``evaluation of individual behavior'' as the primary entry point to reduce psychological burden. Finally, during the data processing stage, anonymous codes replaced real identities, and any potentially identifying information was removed or blurred. As compensation, each participant received 100 RMB upon completion of the interview.

\subsection{Interview Process}

Data were collected through semi-structured interviews. All interviews were conducted online via Tencent Meeting, with each session lasting approximately 70--90 minutes. Interviews were conducted in Chinese and recorded with the participants' permission.

The interview outline was organized around several core modules: (1) Platform usage background: Participants were invited to recall their earliest use of gay dating platforms and their primary motivations, such as meeting friends, dating, seeking partners, or general socializing. (2) Platform use post-relationship formation: Interviews inquired whether participants continued using these platforms after the confirmation of an intimate relationship and whether partners discussed platform usage (e.g., deleting apps or sharing account information). (3) Review of key conflict events: Participants were asked to recall a typical or particularly memorable conflict within a relationship and describe the process in detail, including triggering cues (e.g., online status, profile updates, distance changes, screenshots, or chat logs), interaction styles during the conflict, and how the conflict concluded or was temporarily mitigated. (4) Post-conflict relationship negotiation: The final section explored whether new relationship rules were formed following the conflict (e.g., mandatory app deletion, disclosure of accounts or chat logs, tacit agreements not to discuss platform use, or the formation of open or layered relationship structures). This section focused on the formation of relationship governance settlements and whether these settlements were re-triggered in subsequent interactions.

\subsection{Data Analysis}

All interview recordings were fully transcribed into text following the sessions. Personal identifiers were removed during transcription, and participants were assigned codes (e.g., P1, P2). After transcription, the researcher read the texts multiple times to gain familiarity with the data and identify preliminary themes. The materials were then imported into qualitative analysis software for coding and organization.

Data analysis followed an iterative thematic analysis process \cite{ref8,ref42} consisting of three stages:

(1) Open Coding: The researcher performed line-by-line reading and open coding of all 23 transcripts to identify key concepts mentioned by participants when describing relationship conflicts. This initial pass generated 187 preliminary codes, which were organized into a coding spreadsheet with associated transcript excerpts. Examples of initial codes include ``online status being seen,'' ``profile update discovered,'' ``chat screenshots as evidence,'' and ``deleting then re-downloading apps.'' Each code was accompanied by the participant identifier and the relevant transcript segment to maintain traceability.

(2) Thematic Categorization: Building on preliminary codes, the researcher used cross-case comparisons to aggregate similar codes into higher-level analytical themes. Codes were iteratively grouped, split, and merged through multiple rounds of comparison across transcripts. For example, codes related to ``online status being seen,'' ``distance display noticed,'' and ``profile photo updated'' were grouped under the theme of Visibility Cues, while ``screenshots taken,'' ``chat logs saved,'' and ``profile changes recorded'' were grouped under Evidential Mechanisms. This process yielded 12 intermediate themes, which were further consolidated into three overarching analytical categories: triggering structures, negotiation labor, and governance settlements. Throughout this stage, the researcher paid attention to both convergent patterns across cases and divergent cases that did not fit emerging themes. When participants' accounts diverged from the dominant pattern---for example, participants who reported no conflict over platform use, or who described governance settlements not captured by the initial categories---these cases were examined in detail to assess whether they represented exceptions, boundary conditions, or the need to refine the thematic structure.

(3) Construction of the Relational Process Model: Following thematic categorization, the researcher analyzed the dynamic interaction of these themes within relationships. Cross-case comparisons revealed a recurring cyclical structure in many conflicts: Technical Signal Trigger $\rightarrow$ Conflict Outbreak $\rightarrow$ Negotiated Labor $\rightarrow$ Phased Governance Settlement $\rightarrow$ Re-triggering. Based on this pattern, the study identified several types of relationship governance settlements, such as: Infrastructure Removal (deleting apps/accounts),Transparency Governance (sharing accounts/accessing chat logs), Silence Agreements (ceasing discussion of platform use), and Layered or Open Agreements (redefining relationship boundaries). These settlements do not represent stable endpoints but rather temporary governance structures formed by partners at specific stages. It should be noted that the four settlement types were derived inductively from the data rather than imposed a priori; their boundaries were refined through iterative comparison until each type captured a distinct combination of responses to the governance problems of platform presence, visibility distribution, and loyalty boundaries.

\subsection{Researcher Reflexivity}
This study adopts an interpretive qualitative approach; therefore, the researcher's positionality significantly influenced the research process. During the interviews and analysis, the researcher recognized that participants often sought a balance between "self-justification" and "relational reflection" when describing conflicts. Consequently, the interview design deliberately avoided equating platform use with moral judgment, instead guiding participants to describe specific interaction processes through event retrospection. During the analysis stage, the researcher repeatedly compared narratives across different cases to reduce reliance on a single explanatory framework and ensure that analytical themes emerged from the participants' experiences rather than pre-existing theoretical models. As detailed in Sections 3.1 and 3.2, our sensitizing concepts and sampling boundaries also condition this reflexive stance: we present our themes as situated interpretations whose transferability is bounded by both.

\section{Results}

Our interviews suggest that when gay dating applications (Blued, Aloha, Fanka) continue to exist after an intimate relationship has been formed, they are no longer understood merely as tools for ``meeting new people,'' \cite{ref4,ref23} but gradually transform into a form of relational infrastructure that must be continuously handled and arranged by both partners. Specifically, these platforms continuously provide two types of capabilities that are highly sensitive within intimate relationships. First, they provide visibility cues (such as online status, activity, distance, profile photos, profile updates, etc.), which enable partners to infer whether the other person remains within a ``substitutable social market.'' Second, they provide low-cost and reversible exit/entry mechanisms (apps can be downloaded, logged into, or new accounts can be created at any time), meaning that ``exiting the dating market'' is no longer a one-time commitment but becomes an action that may need to be repeatedly confirmed. Under these conditions, whether a relationship can remain stable often depends not only on emotional compatibility and shared values, but also on whether partners are willing to invest sufficient negotiation labor when platform signals repeatedly appear, in order to temporarily restore a ``workable order of interpretation and rules.''

Through cross-case comparison, we summarize the relevant process as a recurring cycle: technological signals trigger conflict $\rightarrow$ both partners invest negotiation labor such as explanation, control, and rule formation $\rightarrow$ a temporary governance settlement is formed to reduce the frequency of triggers $\rightarrow$ the process may be triggered again under specific conditions by platform signals or external events. The following sections elaborate three key stages of this cycle in sequence: how conflicts are triggered by technological signals (4.1), how negotiation labor unfolds (4.2), and how negotiation eventually sediment into four relatively stable yet highly conditional governance settlements (4.3).

\subsection{Triggering Structure: How Platform Signals Lower the Threshold for Questioning and Reshape the Form of Disputes}

Participants rarely equate ``continuing to use dating applications'' directly with infidelity. However, the multiple visible cues provided by platforms, together with records that can be saved and forwarded, make questioning easier to initiate and make disputes more likely to shift from ``interpretation of meaning'' toward confrontations over ``proof and counter-proof.'' In our materials, we identified four relatively stable triggering pathways: the persistent accessibility of entry points (4.1.1), the evidentiary and shareable nature of usage records (4.1.2), the visibility of attention signals leading to attention accounting (4.1.3), and the combination of distance/location signals with risks of acquaintance recognition and forced outing in the Chinese context (4.1.4). These pathways are not mutually exclusive and may co-occur within a single relationship, but they collectively point to one change: platforms make certain suspicions and dissatisfactions---previously difficult to ``fix as an object''---easier to identify as ``suspicious signals,'' thereby lowering the threshold for initiating questioning.

\subsubsection{The Suspicion of ``Still Using'': Persistent Entry Points Turn Exclusive Commitment into a State that Requires Repeated Confirmation}

When discussing whether using social platforms during an ongoing intimate relationship has a direct causal relationship with infidelity, all participants rejected such a direct equivalence. However, among the 23 participants, 19 mentioned that ``continuing to use'' these platforms would trigger suspicion or anxiety. Platforms thus trigger risk assessments through ``structural possibilities'' rather than actual behavioral facts: conflict starts not from ``what happened,'' but from ``the entry point still exists.''

For example, P2: ``If you still keep the entry point to the dating market, then exclusivity is no longer a default fact. But I think intimate relationships should be exclusive, and now it becomes something that must be explained and reconfirmed.''

Participants repeatedly emphasized not ``he must have done something,'' but ``I cannot be sure whether he might do something,'' which makes trust more easily transformed into an object that must be verified and clarified.

Furthermore, ``continuing to use'' creates an asymmetric information structure: the user can claim ``I didn't do anything,'' while the partner cannot verify that ``nothing happened.'' Multiple participants (P21, P23) described recurring cycles of unverifiable innocence claims. For example, P21:

\begin{quote}
``I was always arguing with my ex about this. Whenever I asked him, he would say he didn't mean anything toward those people (referring to people he chatted with on Blued). But that's just what he says. I think it's unfair to me. We were always arguing about whether using Blued with that intention counts as cheating. And you know what happened later---I was eventually cheated on with someone he met on Blued.''
\end{quote}

This narrative also suggests that when a boundary violation does occur later, the signal of ``the entry point still existed'' may retrospectively be interpreted as an ``earlier sign of risk.''Thus ``continuing to use'' is insufficient to prove that something happened, yet sufficient to reshape how disputes are interpreted, forcing both partners to repeatedly confront: ``Why are you still using it?''

\subsubsection{An ``Evidentiary'' Environment: How Usage Records and Shareability Compress Interpretive Space and Introduce ``Revisiting Old Evidence / Third-Party Judgment''}

If ``continuing to use'' primarily produces suspicion through the persistence of entry points, then screenshots of profiles, online status, and chat fragments further transform the interactional form of disputes. Behaviors that might previously have been negotiated within context are converted into ``evidentiary objects'' that can be repeatedly displayed and circulated across situations. For example, P11:

\begin{quote}
``He kept repeating that he loved me, but at the same time I kept discovering that his Fanka profile kept updating with new photos. How much credibility can that have?''
\end{quote}

Across these accounts, a common interactional pattern emerges: platform-generated traces---profile updates, online duration, status labels---are treated not as ambiguous data points requiring contextual interpretation but as quasi-objective evidence that undermines verbal reassurances. This shift from contextual negotiation to evidence-based confrontation represents a distinctive way in which platform affordances restructure the form of intimate disputes, extending CSCW scholarship on how digital traces reshape accountability \cite{ref55}.

At the same time, evidentiary environments may produce feelings of being judged and experiences of ``being unable to explain clearly.'' For example, P18:

\begin{quote}
``He insisted that the `single' status on my profile meant I was cheating. But actually I had just forgotten to update it. No matter how I explained it, I couldn't make it clear. That made it very difficult.''
\end{quote}

These materials can be stored and re-invoked in future conflicts, making disputes more likely to enter cycles of ``revisiting old evidence.'' Moreover, shareability may expand the conflict from a dyadic negotiation into third-party judgment, further compressing negotiation space. For example, P18:

\begin{quote}
``He sent the screenshots to his friends to judge who was right. That made me very unhappy.''
\end{quote}

\subsubsection{Attention Accounting and ``Newcomer Crisis'': How Online Status and Response Gaps Turn ``Being Valued'' into a Countable Object and May Escalate into Moral Evaluation}

By making attention cues such as online status and activity duration visible, platforms turn the allocation of time and the order of responses into focal points of conflict.

For example, P18: ``He (the participant's ex) can play with whoever he wants, and play however he wants (on the Blued platform), I don't care, but what makes me angry is that I clearly noticed he was online, yet he didn't reply to my messages for almost 10 hours.''

This narrative presents a repeatable chain of inference: online (visible) $\rightarrow$ no reply (visible) $\rightarrow$ lower priority (inferred) $\rightarrow$ relationship value is reduced (inferred) $\rightarrow$ relationship commitment is unreliable (inferred). This chain can be activated without evidence of infidelity, because it revolves around ``relationship ranking'' rather than boundary-crossing facts. In some narratives, attention accounting further overlaps with profile presentation (such as a ``single'' label), and conflict escalates into assessments of integrity and character.

For example, P19 stated: ``I think people should have basic social integrity...I think we have to make this clear. I noticed my partner's Aloha says `single.' I think that's not honest. I think I could accept an open intimate relationship, but it's very hard for me to accept him neglecting me for a long time, and even harder for me to accept him deceiving other boys by saying he's single. I can't continue a relationship with a person who is so immoral.''

Synthesizing across these cases, attention accounting represents a general interactional pattern in which platform-mediated visibility transforms the subjective experience of ``feeling valued'' into a quantifiable and disputable object---a process that extends existing findings on ambient awareness \cite{ref34} by showing how such awareness becomes not merely a background condition but an active resource for relational conflict and moral evaluation.

\subsubsection{Distance/Location and the Compound Trigger of ``Acquaintance Risk'': Loyalty Disputes and External Exposure Risks Occur in Parallel}

In the materials, distance and location cues are not only triggers of distrust within intimate relationships, but also often simultaneously activate external social risks. In the Chinese context, location-based platforms may place users within a field of potential ``recognition by acquaintances.''

\begin{quote}
P17: ``It's not that I'm jealous, it's that I'm afraid something will happen. I already don't use Fanka, but he is still using it. I think this is unsafe for me. We are often together, and if others see him using it, what would they think of me?''
\end{quote}

\begin{quote}
P2: ``I don't want to be forced out. I am extremely disgusted by him (the participant's partner) using Blued, Finka in public spaces.''
\end{quote}

These narratives reveal a key shift that is particularly consequential in the Chinese context, where homosexuality remains socially stigmatized and lacks legal recognition \cite{ref3,ref56}: conflict no longer revolves only around ``loyalty---betrayal,'' but is incorporated into a governance axis of ``visibility---exposure risk.'' Because forced outing can lead to severe consequences---family rupture, workplace discrimination, and social ostracism---the stakes of location-based platform visibility extend far beyond the intimate dyad. 

In these narratives, the object of conflict also involves ``who bears the exposure risk.'' When one partner continues to use a location-based platform, the other may interpret this as outsourcing risk: consequences are borne jointly, but the decision-making power is held unilaterally. 

Therefore, negotiation often shifts toward governance of platform mechanisms and usage contexts, rather than simply demanding exit from the platform. For example, P7: ``I don't force you to delete it, but you at least have to turn off location.''

 Location-based platforms thus produce compound triggers: internal distrust signals and external exposure risks occur in parallel, and disputes often conclude not with a determination of ``whether cheating occurred'' but with rules for ``how to govern visibility.'' This compound trigger structure is distinctive to the Chinese gay community context, where platform use simultaneously involves intimate relationship governance and social exposure risk management---a dual burden absent from the heterosexual relationship contexts that dominate existing research \cite{ref18,ref40}.

\subsection{Negotiation Labor: How Explanation, Control, and Rule Formation Reorganize the Relationship into a ``Workable System''}

After a conflict is triggered, partners do not necessarily break up. Among the 23 participants, 21 reported that the relationship instead entered a phase of intensive negotiation. We summarize this phase as coordination labor within intimate relationships \cite{ref28,ref48}:to keep the relationship operating, partners repeatedly perform three tasks---(1) recoding the meaning of ``use behavior,'' (2) managing access to and visibility on the platform, and (3) turning ambiguous boundaries into executable rules, or turning them into maintainable silence.This labor is not a one-time ``settlement'' but recurs as platform signals reappear, aiming less at full consensus than at a stage-based arrangement that is ``workable enough.''

\subsubsection{Intent Recoding: Pulling ``Use'' Back from the Semantics of Betrayal to Neutral Semantics}

The most common negotiation mechanism is intent explanation: participants attempt to redefine ``still using'' as non-threatening behavior, e.g., browsing, maintaining contacts, or meeting social needs.

For example, P7: ``My ex made things difficult for me before, asking why after being with him I still kept Blued. I remember he said: you already have me, why do you still need Blued? I told him I used Blued because I had many friends on it. If I uninstall it, I will lose many contacts, many friends, social relations.''

These statements appear as defenses on the surface, but interactionally they function more like claims to interpretive authority \cite{ref31}: who gets to define what ``use'' means, thereby preventing the relationship from immediately falling into an interpretive channel of ``betrayal/disloyalty.'' However, intent recoding is fragile: it relies on the partner being willing to accept an ``unverifiable intent.'' In the materials, 16 participants expressed clear dissatisfaction with their partners' explanations, although most also believed that ``an explanation is better than none.''

Therefore, intent explanation is often only the first step of negotiation, frequently followed by more executable governance actions.

\subsubsection{Visibility/Access Control: Negotiation Shifts from ``Tracing Facts'' to ``Managing Trigger Conditions''}
When intent explanations fail to stabilize the relationship, negotiation often becomes materially grounded in controlling visibility. Here, the goal is often not to prove what happened in the past, but to reduce the probability of triggering future conflict: deleting the app, turning off notifications, avoiding opening it in front of the partner, restricting location, hiding photos, and even arranging multiple devices or multiple accounts.

Most participants (16 people) made demands of their partners at this stage, while a small number (4 people) described compensatory strategies in response:

For example, P7: ``You know, I don't want to delete my Blued. I want to keep contact with our community. I think it's very important for my social life. I shouldn't easily cut off my ties with the community. But my partner strongly demanded that I do so, so I bought a second phone and transferred Blued to the second phone. From then on, I didn't use Blued in front of him.''

For example, P2: ``Later I really didn't see him (the participant's partner) continue to scroll Blued. Actually I also didn't know whether he kept scrolling behind my back. But I thought I should have a bit of basic trust and respect for my partner's privacy, so I didn't make too much of it. However, in the end, he still cheated.''

The reversible and multi-platform nature of dating app ecosystems means that controlling visibility on one platform does not eliminate the underlying structural possibility of alternative access---a dynamic that distinguishes these platforms from the social media contexts examined in prior CSCW research \cite{ref21,ref36}.

\subsubsection{Rule Formation and Boundary Declarations: Turning Ambiguous Loyalty into Referencable and Executable Agreements}

In higher-investment negotiation, participants attempt to formulate explicit rules, for example: you can keep it but you can't chat; you can chat but you can't meet; if you meet you must tell first; or restricting usage contexts. Rule formation moves the relationship from ``emotional tacit understanding'' to ``operational agreements,'' enabling both partners to cite boundaries as references in future disputes.

For example, P4: ``In the end what we negotiated was a `standard': you can keep it, but you can't post flirtatious content.''

However, the materials also show that rule formation involves enforcement burdens and platform gray zones.Enforcement is constrained by privacy and by platform affordances (e.g., deleting records, multiple accounts, hiding), so rules may easily slide into ``paper rules.''

For example, P4: ``I already talked about this just now, and I said it very clearly. We (the participant and his partner) said no flirtatious language, only chatting. But whether he was flirtatious or not, how would I know? I don't know his phone password, I don't know his Blued password. Whenever I want to check his records, he gets very unhappy, saying I don't trust him, saying I'm infringing on his privacy.''

When rules are difficult to enforce or too costly to enforce, some relationships shift toward silence agreements, toward more layered/open boundary redefinitions, or directly toward rupture.

\subsubsection{Silence Agreements: Reducing Friction by ``Not Talking,'' but Accumulating Unresolved Tension with High Risk}

When negotiation resources are insufficient, when repeated arguments create fatigue, or when both parties are unwilling to confront boundary differences, a low-cost but high-risk solution is silence: both parties tacitly acknowledge that the platform still exists, but deliberately do not discuss it, do not question it, and do not confirm it.

P23: ``I think this issue has no possibility of being solved, but I'm afraid that if we keep arguing, he (the participant's partner) will break up with me. I'm really afraid of that, really afraid.''

For example, P1: ``When I finally broke up with my ex, I felt really happy. All along it was me trying to communicate, collaborate, and solve problems. We also formulated solutions. He (the participant's partner) didn't follow them. I could only respond with silence, telling him I still cared about him and hoped he would improve. But in the end he still didn't. After that night when he didn't come home all night, I felt everything (referring here to the intimate relationship) was fake. I immediately broke up with him completely.''

\subsection{Four Forms of Stable Governance Settlements: How Negotiation Sediments into Stage-Based, Conditional Arrangements}

After repeated cycles of conflict and negotiation, relationships often sediment into a relatively stable governance settlement. Here, ``stable'' does not mean that the relationship is happier, but rather that both parties form an arrangement that can operate in daily life, reducing trigger frequency, or enabling conflict to be processed under existing rules. We summarize four types of settlements: infrastructure removal, transparency governance, silence agreements, and layered/open agreements. These differ in how they handle three governance problems: platform removal, visibility distribution, and loyalty boundaries.

It should be noted that not all relationships experience all of the above stages. In addition, some participants reported more than one intimate relationship, so the settlement types below are summaries of ``stage-based arrangements within relationship processes,'' rather than classifying each participant into a single category.For example, participant 3 stated that they could not accept their partner's explanation for why they used Blued and Aloha, and after the partner refused their request to delete dating applications, the relationship collapsed completely at that stage. Meanwhile, participants 1, 2, 11, and 23 described their relationships as entering a phase of transparency governance, in which both parties reached different degrees of contracts to defend the relationship's status, but they believed their partners did not comply with these contracts, and the relationships subsequently ended.

\subsubsection{Infrastructure Removal: Trading the Minimum Trigger Probability by Removing Entry Points}
In removal-type settlements, participants tend to view platforms as risk sources that are difficult to govern safely, and therefore choose removal: uninstalling, deleting accounts, or making explicit commitments not to use them anymore. The fragility of such removal echoes Brubaker, Ananny, and Crawford's account of ``leaving'' Grindr, in which departure is not a discrete act but an extended sociotechnical process that platforms make easy to reverse \cite{YA-brubaker2016}.

Participants often described ``deleting it'' as a form of accountability and damage control:

For example, P2: ``If you want to be with me, then delete it. If I delete it, that counts as giving an explanation.''

However, the materials also show that the stability of removal-type settlements can be challenged on multiple fronts. In the Chinese gay community context, deletion carries an additional cost that is often invisible to the demanding partner: because platforms like Blued serve as primary community spaces and identity resources \cite{ref38,ref58}, removing them may sever the user's connection to the broader gay community. Furthermore, because loyalty is bound to a verifiable promise, once re-downloading, alternate accounts, or discovery by a third party occurs, conflict is easily interpreted as a direct betrayal of the promise, thereby escalating.

For example, P10: ``The reason I broke up with one ex was that I found he deceived me. Actually, I didn't have a big problem with whether he used Aloha (during the relationship), or at least it was something we could negotiate. But he said he wouldn't use it, and later I found he secretly used Blued and chatted with others for half a year. You know, being deceived doesn't feel good.''

\subsubsection{Transparency Governance: Reducing Suspicion through Verifiability, but It May Slide into Continuous Checking}

Transparency settlements do not require exiting platforms, but rather incorporate platform use into relationship governance: disclosure, being open to questions, sharing, or ``I'm willing to let you see (P4).'' 

For example, P5: ``I admit that both he and I have social needs. I don't think it's very fair to make him completely disconnect from social apps. More importantly, if he demanded that of me, I also couldn't do it. So we are symmetrical, and this is fair.''

  However, the materials also suggest that transparency is not naturally equivalent to health; its stability depends on a shared boundary of ``checking.'' If transparency becomes continual checking and demands, the platform may transform from contested relational infrastructure into surveillance infrastructure, triggering resistance and concealment.

For example, P18: ``Transparency was useful at first. Later it turned into checking every day, and I couldn't stand it.''

\subsubsection{Silence Agreements: Maintaining Operation by Avoiding the Topic, but at the Cost of Unclear Boundaries.}
Silence settlements usually appear when negotiation resources are insufficient or when repeated conflict produces no results.  Their stabilizing mechanism maintains daily operation by avoiding discussion rather than eliminating risk. Platforms are downgraded into objects ``not allowed to enter the discursive field,'' and the relationship continues to run.

For example, P22: ``Later, I felt that he understood, and I understood, that our relationship was almost over already. We just didn't want to face the collapse of this relationship, even this home. So everyone just pretended not to know.''

  Because it shifts labor into avoidance rather than building a repairable language of rules, silence remains highly dependent on external conditions and grows fragile once a strong trigger occurs.

\subsubsection{Layered/Open Agreements: Institutionalizing Platforms as Part of the Relationship through Boundary Redefinition}
Layered/open settlements are not complete ``laissez-faire,'' but rather a higher-cost form of rule formation \cite{ref25,ref29}. Partners explicitly redefine loyalty boundaries, legalizing platform use and incorporating it into agreements (for example, separating sex and emotion, allowing external interaction but requiring disclosure, or establishing rules for safety and honesty). This often appears after repeated conflict when both parties realize exclusivity rules are difficult to maintain, or when differences in needs cannot be resolved through deletion or transparency. At this point, platforms are no longer treated merely as external threats, but are institutionalized as part of the relationship structure, reducing conflict triggers by reducing concealment.

We report participants' accounts descriptively and make no normative judgment about the relational arrangements they describe. For example, P19: ``My current partner and I are completely equal. Sometimes, when I see someone I like on Blued, I even recommend them to him. He doesn't think there's anything wrong with this, because I also don't demand that he not have dating apps, or even third-party involvement.''

The materials also suggest that the conditional stability of layered settlements relies on symmetry and clarity. If rules are asymmetric, or if openness is nominal while concealment persists, distrust problems may still arise. Therefore, layered/open settlements are not the ``most free'' settlements, but the ``most governance-demanding'' settlements. 

Synthesizing the above analysis, we understand the influence of dating applications on intimate relationships as a restructuring of interactional form \cite{ref4,ref23}. Platforms continuously emit suspicious signals, making questioning easier to initiate. Partners must invest negotiation labor, organizing ``intent explanation—visibility control—rule formation/silence/boundary redefinition'' into stage-based arrangements that can keep the relationship operating. These arrangements ultimately sediment into four settlements, yet each settlement is conditionally stable due to platforms' reversibility, visibility, and evidentiary capacities.  This empirical pattern extends existing findings in CSCW research on technology and intimate relationships \cite{ref18,ref21,ref40}: whereas prior work has primarily documented how digital platforms trigger emotional responses such as jealousy or anxiety, our data reveal that the relational consequences of platform persistence are better understood as ongoing processes of interactional restructuring—processes in which partners actively negotiate, contest, and provisionally stabilize the meaning and governance of platform presence within their relationships.

\section{Discussion}
In this section, we discuss the contributions and implications of our findings, reflect on the limitations of the study, and propose directions for future research.

\subsection{Summary of Findings}

Taken together, our findings show that the suspicion--negotiation--settlement cycle and the four governance settlements it produces never resolve into permanent stability: the reversible entry mechanisms, continuous visibility cues, and evidentiary capacities of platforms keep the conditions for re-triggering perpetually present. Rather than rehearse these results again, we treat this irreducible instability as the analytic point of departure for the discussion that follows. In the remainder of this section we move from describing what couples do to theorizing why platform-mediated intimacy resists closure, situating our account within ongoing conversations on relational maintenance, infrastructural visibility, and the governance of intimate life.

\subsection{Empirical Contributions}
Our study makes three empirical contributions that extend existing CSCW and HCI scholarship on digital technology and intimate relationships. First, existing research has extensively documented how social media platforms trigger jealousy, surveillance behaviors, and trust issues in intimate relationships \cite{ref18,ref21,ref36,ref40}. Our findings build on this work by revealing a more complex process: platform signals do not merely provoke emotional reactions but prompt partners to engage in sustained relational negotiation---through explanation, visibility management, rule formation, or silence---that continuously reshapes how intimate partnerships are structured and governed. By tracing the full cycle from triggering to negotiation to settlement, we empirically demonstrate how platform affordances---visibility, persistence, reversibility, and evidentiation---alter the interactional conditions of intimate relationships at a structural level, rather than merely amplifying pre-existing emotional tendencies \cite{ref10,ref55}. 

Second, we extend the application of the relational infrastructure concept \cite{ref45,ref49} to the domain of intimate relationships by showing how dating applications, once embedded within an established relationship, function not as neutral tools but as persistent sociotechnical structures that continuously generate governance demands. Our empirical data illustrate that platform presence creates ongoing requirements for maintenance, negotiation, and repair---what we term negotiation labor \cite{ref28,ref30,ref48}. Rather than re-defining the construct here (see Sections 1 and 2.2), we position its theoretical contribution against the CSCW repair-work literature \cite{YA-jackson2014,YA-henke1999,YA-graham2007}: like repair work, negotiation labor is continuous, frequently unacknowledged, unevenly distributed, and visible mainly at breakdowns; unlike repair work, it is generative rather than restorative---producing new terms, rules, and boundaries instead of returning to a prior state---and it targets the relational order itself rather than a material artifact. Negotiation labor thus reframes repair as a relational, forward-looking accomplishment. 

Third, we extend the literature on LGBTQ+ intimate relationships and technology \cite{ref25,ref33,ref44} by providing empirical evidence of how the social and cultural context of Chinese gay communities shapes the negotiation of platform use within relationships. In a context where dating platforms serve simultaneously as dating tools, community spaces, and identity resources \cite{ref11,ref38,ref58}, the common assumption that deleting a dating application is a straightforward marker of commitment becomes untenable.  This finding challenges both the normative expectation that platforms should be abandoned after relationship formation and the assumption that continued use necessarily signals disloyalty.

\subsection{The Specificity of the Chinese Gay Community Context}

Our findings reveal that the relational dynamics surrounding dating app use among Chinese gay men cannot be fully understood without attending to the distinctive social functions these platforms serve in this context. Unlike in many Western settings where dating apps operate alongside a relatively visible LGBTQ+ social infrastructure (community centers, pride events, legally recognized partnerships), dating platforms such as Blued, Aloha, and Fanka function in China not only as tools for romantic or sexual connection but simultaneously as community spaces, identity resources, and social support networks \cite{ref11,ref38,ref58}. This multifunctionality means that the act of ``deleting the app''---often treated in existing literature as a straightforward marker of relationship commitment---carries a fundamentally different meaning in this context: it may also entail severing ties with the broader gay community, losing access to identity-affirming spaces, and disconnecting from social support systems that are difficult to replace through other channels \cite{ref37}. 

Furthermore, the location-based features of these platforms introduce risks that are specific to the Chinese social context. Because homosexuality in China occupies a legal and cultural gray zone \cite{ref3,ref56}, proximity-based visibility on dating apps creates the possibility of acquaintance recognition and forced outing---risks with potentially severe social consequences including family rupture, workplace discrimination, and social ostracism.  Partners must negotiate not only ``are you faithful to me'' but also ``are you exposing us to the risk of being identified.'' This compound governance problem---where intimate relationship management and social risk management are inseparable---represents a distinctive feature of platform use in this context that existing CSCW research on technology and intimate relationships, largely conducted with heterosexual couples in Western settings \cite{ref18,ref40,ref53}, has not addressed.

These dynamics also invite a closer engagement with literature that theorizes how privacy and boundaries are managed within intimate relationships through means other than explicit declaration. Communication Privacy Management theory frames intimate partners as co-owners of private information who must continually coordinate boundary rules, and it foregrounds the ``boundary turbulence'' that erupts when those rules are violated or left implicit \cite{YA-petronio2002}. Our participants' reluctance to articulate platform expectations aligns with research on privacy silence and topic avoidance, which shows that partners frequently choose not to raise sensitive matters in order to protect the relationship, preserve face, and avoid conflict \cite{YA-caughlin2004}. For Chinese gay men, such silence is doubly motivated: it guards the relationship and it manages the additional risk of exposure within a stigmatizing social environment \cite{ref3,ref56}. Crucially, much of this negotiation proceeds non-verbally. Rather than stating rules, partners communicate expectations through monitored behavior, the visible (or strategically invisible) presence of an app icon, and the symbolic act of deletion itself---practices consonant with accounts of tacit boundary coordination and topic avoidance in intimacy \cite{YA-petronio2002,YA-caughlin2004}. Reading platform conduct as a medium of unspoken negotiation helps explain why ``deleting the app'' functions less as a discrete decision than as an ongoing, ambiguous communicative gesture whose meaning both partners must interpret without ever speaking it aloud.

\subsection{Design Implications}
Our findings suggest several design implications for dating platform developers. First, the triggering structure we identified reveals that many conflicts are activated not by actual boundary-crossing behavior but by structural signals that platforms make visible by default. Features such as real-time online status, precise distance indicators, and activity duration displays create conditions for continuous monitoring and attention accounting. Platform designers should consider providing users with more granular control over which signals are visible to others, including the option to selectively disable status indicators without fully withdrawing from the platform \cite{ref5,ref35}. We offer these recommendations cautiously, however, and resist framing them as straightforward fixes. The same indicators that enable surveillance also do consequential work for users: online-status and distance signals support partner discovery and the coordination of meetings, allow people to gauge safety and authenticity before engaging, and help sustain a felt sense of an active community---functions that participants themselves valued even as they chafed at the monitoring such signals afforded. Removing or hiding these signals therefore involves a trade-off rather than a clear improvement, and may shift burdens elsewhere. Granular visibility controls likewise carry unintended consequences: as our data on the interpretive economy of signals suggests, the very act of selectively concealing one's status can itself become a suspicious signal, read by a partner as evidence of something to hide. Design here cannot resolve the underlying relational tensions; at best it can redistribute the interpretive labor and, ideally, avoid amplifying the default visibility that makes ordinary platform use legible as accountable activity in the first place.

Second, the evidentiary dynamics we documented---in which screenshots, saved profiles, and forwarded records transform disputes from contextual negotiations into confrontations over evidence---suggest that platforms should consider how their features contribute to the objectification and decontextualization of interpersonal interactions. While users cannot be prevented from taking screenshots, design choices such as ephemeral content, contextualized activity logs, or notification when profiles are being viewed could help restore some interpretive space within relational disputes. We again note that such measures are not without tension: notifying users when their profile is viewed could equally intensify monitoring rather than alleviate it, and ephemerality can frustrate users who rely on records for safety or accountability. These features should thus be understood as reallocating interpretive space, not neutrally expanding it.

Third, the finding that dating platforms serve multiple functions for Chinese gay men---not only as dating tools but as community spaces and identity resources---suggests that platform design should accommodate these diverse use cases. Features that allow users to signal relationship status in nuanced ways, or that separate community and social functions from active dating features, could reduce the assumption that any platform use implies romantic or sexual availability. To be clear, by ``separating'' community and dating functions we mean offering distinct modes or contexts \emph{within} a single application---for example, a community or social browsing mode whose use does not register as dating-related activity---rather than advocating that these functions be split across separate platform infrastructures. Fragmenting them into different apps would risk reintroducing the very legibility problem we describe, since being present on a dedicated dating platform would once more carry an unambiguous interpretation. The aim is to loosen the equation between presence and availability, not to build new walls that re-fix it. Across all three directions, we therefore frame our recommendations not as fixes but as \emph{design trade-offs} in the CSCW tradition of socio-technical tension analysis: every signal that can be surveilled also does connective, safety, or identity work; every concealment option can itself become a signal; and any redistribution of visibility redistributes interpretive labor between partners rather than eliminating it. Design in this space chooses among tensions; it does not resolve them.

\subsection{Limitations and Future Work}

This study has several limitations that should be acknowledged. First, our sample of 23 participants, while providing rich qualitative data, cannot capture the full diversity of experiences among Chinese gay men. The reliance on online recruitment and snowball sampling may have excluded individuals who are less connected to LGBTQ+ online communities or who have limited digital access. Future research should explore strategies for reaching more diverse populations.

Second, our study relied on retrospective accounts of relationship conflicts, inevitably shaped by memory and emotional processing. Real-time or diary-based methods could complement interview data by capturing conflicts and negotiations as they unfold.

Third, our analysis focused on individual participants' accounts rather than dyadic data. Crucially, we interviewed only one partner per relationship, which constitutes a dyadic-perspective limitation: our data reflect a single situated account of each arrangement rather than the relationship as a jointly enacted system. As a result, we cannot capture the interactional dynamics that unfold between partners, nor the mutual perception through which each reads, anticipates, and responds to the other. What one partner narrates as a settled understanding may be experienced by the other as unresolved or coercive, and the negotiation labor we describe is necessarily seen from one vantage point only. Interviewing both partners within the same relationship would provide richer insight into how negotiation labor is distributed, how governance settlements are co-constructed, and how each partner's understanding of the arrangement may differ.

Fourth, our study was conducted within the specific socio-cultural context of contemporary China, where same-sex relationships lack legal recognition and LGBTQ+ expression is subject to regulatory uncertainty \cite{ref3,ref56}. The extent to which our findings generalize to other cultural and legal contexts requires further investigation. Comparative studies across different settings could illuminate how legal recognition, social acceptance, and platform governance regimes shape the relational dynamics we identified.

A further conceptual point warrants emphasis, as it bears on how our findings should be read. One might assume that dating apps become redundant once a relationship forms and the search for a partner ends. Our data suggest the opposite: the apps do not disappear from a couple's life but persist as socio-technical infrastructure that continues to reshape how conflict is triggered, negotiated, and governed. Lingering profiles, recommendation feeds, read receipts, and the latent possibility of re-encountering alternatives remain materially present and continue to structure suspicion, accountability, and trust long after the relationship is established. Consequently, stability is not achieved by exiting the platform but must be continually produced through ongoing negotiation labor. The settlements partners reach are therefore better understood as a temporary equilibrium---provisional, renegotiated as platform features and circumstances change---rather than a permanent resolution. This reframing is consistent with recent scholarship on platformized intimacy and relational labor, which shows that intimate ties are increasingly mediated by persistent platform architectures and sustained through continuous, often invisible, relational work \cite{YA-bandinelli2022,YA-pym2021,YA-duguay2020}.

Finally, our conceptual framework---the triggering--negotiation--settlement cycle---was developed inductively from our data. Future research could test and refine this framework through larger-scale studies, longitudinal designs, or application to other relational contexts in which persistent digital infrastructures reshape the conditions of intimate interaction.

Despite these limitations, this study contributes to understanding how platforms restructure the interactional conditions of intimate relationships, inviting both researchers and designers to attend more carefully to the relational consequences of platform persistence.

[Note on revision: the standalone ``Limitations and Future Work'' section that previously followed the Discussion duplicated Section 5.5 verbatim and has been removed; all limitations material now appears once, in Section 5.5.]

\section{Conclusion}
This study investigated how the persistent presence of gay dating applications shapes intimate relationships among 23 Chinese gay men. Through semi-structured interviews and thematic analysis, we identified a recurring cyclical process: platform signals trigger suspicion and conflict, partners invest negotiation labor to manage these conflicts, and negotiations sediment into governance settlements that provide conditional but never permanent stability.

Our central empirical contribution is demonstrating how dating applications function as relational infrastructure—persistent sociotechnical structures that, once embedded within an established relationship, continuously generate governance demands. We identified four forms of governance settlements (infrastructure removal, transparency governance, silence agreements, and layered/open agreements), each representing a distinct arrangement for managing the problems of platform presence, visibility distribution, and loyalty boundaries. The conditional stability of all four settlements—arising from platforms' reversible entry mechanisms, continuous visibility cues, and evidentiary capacities—demonstrates that platforms do not directly cause relational harm but fundamentally restructure the conditions under which conflicts are triggered, disputes are conducted, and temporary stability is achieved.

These findings have implications for both CSCW research and platform design. For researchers, we offer an empirical extension of existing work on digital technology and intimate relationships—through the triggering–negotiation–settlement cycle—that moves beyond individual psychological responses to examine the interactional dynamics through which couples manage platform-related tensions. For designers, we highlight how default visibility features, evidentiary affordances, and the conflation of dating and community functions create structural conditions for relational conflict that could be mitigated through more granular user controls. We hope this study encourages further investigation into the relational consequences of persistent digital infrastructures, particularly in non-Western contexts where platform use intersects with distinctive social, cultural, and regulatory conditions.

\bibliographystyle{ACM-Reference-Format}
\bibliography{references}

\appendix

\section{Participant Background Information}

\begin{table}[h]
\centering
\caption{Participant Background Information}
\begin{tabular}{@{}llllll@{}}
\toprule
ID & U/R & Education & Profession & Rel. Status & Dating Apps \\
\midrule
 & U & Bachelor & Student & Y & Blued, Aloha \\
 & U & High School & Singer & Y & Blued, Aloha, Fanka \\
 & U & Bachelor & Student & N & Blued, Aloha \\
 & R & High School & Chef & N & Aloha, Fanka \\
 & R & Bachelor & Math Teacher & N & Blued, Aloha, Fanka \\
 & U & Bachelor & Reporter & Y & Blued \\
 & R & Bachelor & None & Y & Blued, Aloha, Fanka \\
 & U & Master & Nurse & Y & Blued \\
 & R & High School & Village Secretary & N & Aloha, Fanka \\
 & R & Junior High School & Farmer & N & Blued \\
 & U & PhD & Doctor & Y & Blued, Aloha, Fanka \\
 & U & Bachelor & Student & Y & Blued, Fanka \\
 & R & Bachelor & Courier & Y & Aloha, Fanka \\
 & R & Bachelor & Student & Y & Blued, Aloha, Fanka \\
 & R & Primary School & Bartender & N & Aloha, Fanka \\
 & U & High School & Student & Y & Blued, Aloha \\
 & U & High School & Student & Y & Blued \\
 & U & Junior High School & Labour & N & Blued, Aloha \\
 & R & Junior High School & Labour & Y & Blued, Aloha, Fanka \\
 & R & Bachelor & Nurse & Y & Blued, Aloha, Fanka \\
 & U & Bachelor & Student & N & Fanka \\
 & R & Primary School & Farmer & Y & Blued, Aloha, Fanka \\
 & U & Bachelor & Editor & N & Blued \\
\bottomrule
\end{tabular}
\end{table}

\section{Semi-Structured Interview Guide}

\begin{description}

\item[A0 Ethics and Opening]\hfill
\begin{itemize}
  \item \textbf{A0-1.} \emph{Main Question:} Do you consent to audio recording? The recording will only be used for transcription and analysis and will be de-identified afterward.
  \emph{Follow-up Prompts:} You may pause/skip/terminate the interview at any time; any sensitive details can be described in a blurred or generalized manner.
  \emph{Analytical Mapping:} Ethics, right to withdraw, risk management.
  \item \textbf{A0-2.} \emph{Main Question:} Before we begin, how would you like me to address you? Are there any sensitive topics you prefer not to be asked about?
  \emph{Follow-up Prompts:} Coming out / work / family / sexual behavior etc.\ can be skipped.
  \emph{Analytical Mapping:} Participant comfort, boundary setting.
\end{itemize}

\item[A1 Background and Relationship Context]\hfill
\begin{itemize}
  \item \textbf{A1-1.} \emph{Main Question:} When did you first start using (Blued/Aloha/Fanka/others)? What did you mainly use it for?
  \emph{Follow-up Prompts:} Meeting people / dating / hookups / making friends / browsing updates / community connection / passing time, etc.
  \emph{Analytical Mapping:} Platform use purposes, diversity of uses.
  \item \textbf{A1-2.} \emph{Main Question:} Would you be willing to talk about one intimate relationship you are most familiar with or consider representative? Approximately how long did it last?
  \emph{Follow-up Prompts:} How the relationship was confirmed; whether it was long-distance; whether you lived together.
  \emph{Analytical Mapping:} Relationship context, relationship stage.
  \item \textbf{A1-3.} \emph{Main Question:} After the relationship was confirmed, did your or your partner's use of the app change?
  \emph{Follow-up Prompts:} Delete / not delete; frequency; whether usage was public; whether usage shifted to other platforms.
  \emph{Analytical Mapping:} Persistence of entry points (4.1.1), visibility changes.
  \item \textbf{A1-4.} \emph{Main Question:} Were your expectations regarding ``exclusivity/openess'' aligned at the time? Did you ever discuss boundaries?
  \emph{Follow-up Prompts:} Reasons for not discussing: default assumptions, avoiding conflict, considering it unnecessary, etc.
  \emph{Analytical Mapping:} Boundary consensus/lack thereof, preconditions for negotiation.
\end{itemize}

\item[A2 Key Narrative: Reconstructing a Conflict Cycle (Recommended Core Question)]\hfill
\begin{itemize}
  \item \textbf{A2-1.} \emph{Main Question:} Recall the most typical or intense conflict: how did it initially start? What signals did you notice?
  \emph{Follow-up Prompts:} Be specific: online status, profile updates, distance, screenshots, chat fragments, status labeled ``single'', etc.
  \emph{Analytical Mapping:} Overall entry point of triggering structure (4.1).
  \item \textbf{A2-2.} \emph{Main Question:} What was your first reaction at that moment? What did you do?
  \emph{Follow-up Prompts:} Directly ask, endure silently, investigate, screenshot, check records, ask friends for judgment, remain silent, etc.
  \emph{Analytical Mapping:} Types of negotiation labor (4.2).
  \item \textbf{A2-3.} \emph{Main Question:} How did the other person respond at the time? What explanations did they provide? Why did you accept or reject them?
  \emph{Follow-up Prompts:} ``Just chatting / browsing out of boredom / friends are on the platform / community contact / habitually didn't delete it'', etc.
  \emph{Analytical Mapping:} Intent re-coding (4.2.1), contestation of interpretive authority (4.1.1).
  \item \textbf{A2-4.} \emph{Main Question:} What arrangements did you eventually make that allowed the relationship to ``temporarily continue''?
  \emph{Follow-up Prompts:} Deleting the app, sharing/transparency, setting rules, turning off location, not using it in front of each other, avoiding discussion, open agreements, etc.
  \emph{Analytical Mapping:} Negotiation pathways prior to governance settlement (4.2 $\rightarrow$ 4.3).
  \item \textbf{A2-5.} \emph{Main Question:} Did a ``re-trigger'' occur afterward? What caused the conflict to emerge again?
  \emph{Follow-up Prompts:} Reappearance of platform signals, external events, third-party intervention, old evidence being revisited.
  \emph{Analytical Mapping:} Cyclical structure, conditional stability (overall framework).
\end{itemize}

\item[A3 Triggering Structure: Persistence of Entry Points]\hfill
\begin{itemize}
  \item \textbf{A3-1.} \emph{Main Question:} In your view, what does it mean if someone continues to keep or use the app after confirming a relationship? What is the most painful aspect?
  \emph{Follow-up Prompts:} ``The entry point still exists'' vs ``behavior crossing the line''; which matters more to you?
  \emph{Analytical Mapping:} Entry point persistence, exclusivity commitments requiring repeated confirmation.
  \item \textbf{A3-2.} \emph{Main Question:} Would you equate ``continued use'' directly with cheating? If not, what does it trigger instead?
  \emph{Follow-up Prompts:} Suspicion / anxiety / risk evaluation / suspended boundaries / trust being tested.
  \emph{Analytical Mapping:} Structural possibility triggering rather than factual events.
  \item \textbf{A3-3.} \emph{Main Question:} In these conflicts, what is harder: proving what the other person did, or proving what they did not do?
  \emph{Follow-up Prompts:} Have you attempted to request chat records/passwords? Why not / why did the other person refuse?
  \emph{Analytical Mapping:} Non-falsifiability structure, information asymmetry.
  \item \textbf{A3-4.} \emph{Main Question:} During arguments, do discussions revolve more around ``facts'' or around ``intentions/meanings''?
  \emph{Follow-up Prompts:} ``Does using it count as disloyalty / crossing boundaries?'' Who has the authority to define this?
  \emph{Analytical Mapping:} Contestation of interpretive/definitional authority (4.1.1).
\end{itemize}

\item[A4 Triggering Structure: Evidentiation and Shareability]\hfill
\begin{itemize}
  \item \textbf{A4-1.} \emph{Main Question:} Have you or your partner ever taken screenshots or saved profiles, online status, or chat fragments? Why save them?
  \emph{Follow-up Prompts:} Saved for self-justification / accusation / keeping records / using later.
  \emph{Analytical Mapping:} Objectification of evidence, authority of evidence.
  \item \textbf{A4-2.} \emph{Main Question:} What role did screenshots or records play in arguments? Did they help resolve issues or escalate confrontation?
  \emph{Follow-up Prompts:} Situations where ``it cannot be explained clearly'' or ``the more you explain, the more it looks like an excuse''.
  \emph{Analytical Mapping:} Context compression, reduction of interpretive space.
  \item \textbf{A4-3.} \emph{Main Question:} Did ``bringing up old accounts'' occur, where old screenshots were repeatedly brought up again? What did that mean to you?
  \emph{Follow-up Prompts:} How old evidence was re-mobilized; whether it prevented the relationship from resetting.
  \emph{Analytical Mapping:} Evidence circulation / continuous auditing.
  \item \textbf{A4-4.} \emph{Main Question:} Did you ever forward screenshots or records to friends or third parties to ask for judgment? Why?
  \emph{Follow-up Prompts:} Feelings of being observed or judged; changes after third-party involvement.
  \emph{Analytical Mapping:} Third-party adjudication, compression of negotiation space.
\end{itemize}

\item[A5 Triggering Structure: Attention Signals]\hfill
\begin{itemize}
  \item \textbf{A5-1.} \emph{Main Question:} Have you experienced situations where you saw the other person online but they did not reply? How did you interpret that at the time?
  \emph{Follow-up Prompts:} Online = having attention; not replying = attention directed elsewhere.
  \emph{Analytical Mapping:} Attention accounting, prioritization framework.
  \item \textbf{A5-2.} \emph{Main Question:} How is this type of conflict different from conflicts about ``whether cheating occurred''?
  \emph{Follow-up Prompts:} More about ``being ignored / low priority'' or ``suspected boundary violation''?
  \emph{Analytical Mapping:} Shift from suspicion of violation to contestation of relational ranking.
  \item \textbf{A5-3.} \emph{Main Question:} Do you believe the other person has an ``obligation to explain'' (at least explaining why they were online but did not respond)?
  \emph{Follow-up Prompts:} Is the explanation meant to provide information, or to repair relational order?
  \emph{Analytical Mapping:} Institutionalization of explanatory obligation, reparative labor.
  \item \textbf{A5-4.} \emph{Main Question:} Would you interpret online presentation (e.g., a ``single'' label) as an issue of honesty or character?
  \emph{Follow-up Prompts:} Acceptable to be open but not acceptable to deceive? Why?
  \emph{Analytical Mapping:} Pathway of moral escalation (to be interpreted cautiously).
\end{itemize}

\item[A6 Trigger: Location/Familiar Network]\hfill
\begin{itemize}
  \item \textbf{A6-1.} \emph{Main Question:} Do you care about location/distance features? Is the concern about loyalty risk, or about being seen/being outed by acquaintances?
  \emph{Follow-up Prompts:} How do the two risks combine? Which scenarios feel most dangerous?
  \emph{Analytical Mapping:} Compound trigger: intimate conflict + external exposure.
  \item \textbf{A6-2.} \emph{Main Question:} Do you dislike it when your partner opens or uses the app in public places? Why?
  \emph{Follow-up Prompts:} Imagining colleagues/family/acquaintances encountering it; possible consequences.
  \emph{Analytical Mapping:} Safety boundaries, situational risk.
  \item \textbf{A6-3.} \emph{Main Question:} If acquaintances encounter the profile, who bears the consequences? Do you think decision power and consequence bearing are symmetrical?
  \emph{Follow-up Prompts:} Situations where ``both people need to explain themselves''.
  \emph{Analytical Mapping:} Asymmetry between decision power and consequence bearing.
  \item \textbf{A6-4.} \emph{Main Question:} Did you negotiate specific parameters or situational rules (turning off location, reducing radius, prohibiting usage in certain contexts)?
  \emph{Follow-up Prompts:} Which were acceptable and which were unacceptable?
  \emph{Analytical Mapping:} From meaning interpretation to risk governance (parameterized governance).
\end{itemize}

\item[A7 Negotiation Labor: Intent Re-coding]\hfill
\begin{itemize}
  \item \textbf{A7-1.} \emph{Main Question:} How does the other person usually explain continued use? Which explanations are you more likely to accept?
  \emph{Follow-up Prompts:} ``Friends are on the platform / community contact / browsing out of boredom / habitually didn't delete it'', etc.
  \emph{Analytical Mapping:} Intent re-coding, reconstruction of meaning frameworks.
  \item \textbf{A7-2.} \emph{Main Question:} Do you need ``an explanation'' more, or ``verifiable actions''? When do you shift from the former to the latter?
  \emph{Follow-up Prompts:} What moment made you decide the app must be deleted / transparency must be implemented / rules must be established?
  \emph{Analytical Mapping:} Conditions for shifting from explanation to control/rule formalization.
  \item \textbf{A7-3.} \emph{Main Question:} In your view, what is most central in these disputes: ``trust'', ``boundaries'', or ``respect''?
  \emph{Follow-up Prompts:} How do you define these terms?
  \emph{Analytical Mapping:} Participant ontological understanding (avoiding author substitution).
\end{itemize}

\item[A8 Negotiation Labor: Visibility/Access Control]\hfill
\begin{itemize}
  \item \textbf{A8-1.} \emph{Main Question:} Have you ever made specific requests: delete the app / don't use it in front of me / turn off notifications / turn off location / don't update the profile, etc.?
  \emph{Follow-up Prompts:} Which requests were accepted or rejected by the other person? What reasons were given?
  \emph{Analytical Mapping:} Visibility control, trigger condition management.
  \item \textbf{A8-2.} \emph{Main Question:} Did the other person adopt compensatory strategies: second phone, secondary account, new account, switching platforms, deleting records?
  \emph{Follow-up Prompts:} When did you discover it? How did you interpret ``hiding'' after discovering it?
  \emph{Analytical Mapping:} Double-edged effect of control: short-term stability / long-term asymmetry.
  \item \textbf{A8-3.} \emph{Main Question:} Did these controls lead to ``fewer arguments'' or ``stronger suspicion''? Why?
  \emph{Follow-up Prompts:} Did this make you check more frequently or feel more inclined to give up the relationship?
  \emph{Analytical Mapping:} Changes in trigger frequency, rising relational costs.
\end{itemize}

\item[A9 Negotiation Labor: Rule Formalization]\hfill
\begin{itemize}
  \item \textbf{A9-1.} \emph{Main Question:} Did you ever establish explicit rules? What were they?
  \emph{Follow-up Prompts:} Can keep the app but cannot chat; can chat but cannot meet; meetings must be disclosed beforehand; profile labeling, etc.
  \emph{Analytical Mapping:} Rule formalization / boundary declaration.
  \item \textbf{A9-2.} \emph{Main Question:} What was the hardest part to implement?
  \emph{Follow-up Prompts:} Privacy boundaries, unverifiability, platform grey zones (deleted records/multiple accounts).
  \emph{Analytical Mapping:} Enforcement costs, grey-zone governance.
  \item \textbf{A9-3.} \emph{Main Question:} When you cannot verify whether the other person follows the rules, what do you do?
  \emph{Follow-up Prompts:} Checking, renegotiation, silence, shifting to openness, ending the relationship.
  \emph{Analytical Mapping:} Path divergence after rule breakdown.
\end{itemize}

\item[A10 Negotiation Labor: Silent Pact]\hfill
\begin{itemize}
  \item \textbf{A10-1.} \emph{Main Question:} Did you experience a stage where ``both people knew but did not talk about it''? Why choose not to talk about it?
  \emph{Follow-up Prompts:} Fatigue, fear of breakup, believing there is no solution, maintaining everyday life.
  \emph{Analytical Mapping:} Silence as a governance strategy.
  \item \textbf{A10-2.} \emph{Main Question:} Did silence bring relief or accumulation of tension? Under what conditions would silence break?
  \emph{Follow-up Prompts:} Strong triggers: not returning home overnight, discovery by a third party, screenshot evidence, etc.
  \emph{Analytical Mapping:} Conditional stability, explosion threshold.
\end{itemize}

\item[A11 Governance Settlement: Overall Question]\hfill
\begin{itemize}
  \item \textbf{A11-1.} \emph{Main Question:} Returning to this relationship: what kind of ``arrangement that allowed the relationship to function'' eventually formed? (Multiple selections/possible transitions)
  \emph{Follow-up Prompts:} Did the relationship shift from one arrangement to another during the process?
  \emph{Analytical Mapping:} Settlements are non-exclusive, processual sedimentation.
  \item \textbf{A11-2.} \emph{Main Question:} Did this arrangement reduce conflict? Did it reduce frequency, intensity, or simply make it ``less quarrelsome but more distant''?
  \emph{Follow-up Prompts:} How do you evaluate this kind of ``stability''?
  \emph{Analytical Mapping:} Stability $\neq$ happiness (operational formulation).
\end{itemize}

\item[A12 Settlement Type: Infrastructure Removal]\hfill
\begin{itemize}
  \item \textbf{A12-1.} \emph{Main Question:} Did you or your partner choose to uninstall, delete the account, or explicitly promise not to use the app again?
  \emph{Follow-up Prompts:} Was this ``giving an account/stopping losses'' or ``control''?
  \emph{Analytical Mapping:} Removal-type settlement: eliminating the entry point.
  \item \textbf{A12-2.} \emph{Main Question:} Under what circumstances does this settlement become unstable?
  \emph{Follow-up Prompts:} Reinstallation, secondary accounts, discovery by third parties, use on another device.
  \emph{Analytical Mapping:} Instability conditions of removal settlement.
\end{itemize}

\item[A13 Settlement Type: Transparency Governance]\hfill
\begin{itemize}
  \item \textbf{A13-1.} \emph{Main Question:} Did you practice notification/sharing/allowing inspection? How did this change suspicion?
  \emph{Follow-up Prompts:} Did verifiability reduce questioning, or trigger more checking?
  \emph{Analytical Mapping:} Transparency governance: attempt at information symmetry.
  \item \textbf{A13-2.} \emph{Main Question:} When does transparency become surveillance? Did you discuss ``inspection boundaries''?
  \emph{Follow-up Prompts:} Who can see what, for how long, and how to stop?
  \emph{Analytical Mapping:} Instability conditions of transparency: surveillance and resistance.
\end{itemize}

\item[A14 Settlement Type: Silent Settlement]\hfill
\begin{itemize}
  \item \textbf{A14-1.} \emph{Main Question:} Did you enter a long-term ``default existence'' state where the issue remained unspoken? What did that mean to you?
  \emph{Follow-up Prompts:} Maintenance / delay / compensation; did you feel the relationship was approaching its end?
  \emph{Analytical Mapping:} Silent settlement: maintaining functionality through avoidance.
  \item \textbf{A14-2.} \emph{Main Question:} What triggers do you fear most that might collapse this silence?
  \emph{Follow-up Prompts:} Types of strong trigger events and their consequences.
  \emph{Analytical Mapping:} Instability conditions of silent settlement.
\end{itemize}

\item[A15 Settlement Type: Layered/Open Agreements]\hfill
\begin{itemize}
  \item \textbf{A15-1.} \emph{Main Question:} Did you experience a redefinition of boundaries: allowing certain external interactions but requiring notification/safety/honesty?
  \emph{Follow-up Prompts:} Separation of sex/emotion; symmetry; rule language.
  \emph{Analytical Mapping:} Layered/open arrangement: boundary redefinition.
  \item \textbf{A15-2.} \emph{Main Question:} What is most difficult in layered/open arrangements: symmetry, clarity, or enforcement?
  \emph{Follow-up Prompts:} How does the situation of ``verbally open but actually concealed'' occur?
  \emph{Analytical Mapping:} Instability conditions of layering: asymmetry and concealment.
\end{itemize}

\item[A16 Counterfactual and Design Implications]\hfill
\begin{itemize}
  \item \textbf{A16-1.} \emph{Main Question:} If you could redesign these apps, which features most easily trigger conflict? How would you change them?
  \emph{Follow-up Prompts:} Online visibility, activity status, screenshots, location, status labels, etc.
  \emph{Analytical Mapping:} Mechanisms and design implications.
\end{itemize}

\end{description}

\end{document}